\documentclass[aps,pra,preprint,groupaddress]{revtex4}

\usepackage[centertags]{amsmath}
\usepackage{graphicx}
\usepackage{mathrsfs}
\usepackage{color}

\begin{document}
\title{Dynamics of the modulation instability spectrum in optical fibers with oscillating dispersion}

\author{M. Droques}
   \affiliation{Universit\'e Lille 1, Laboratoire PhLAM, IRCICA, 59655 Villeneuve d'Ascq, France}
\author{A. Kudlinski}
\email[Email address~:~]{alexandre.kudlinski@univ-lille1.fr} \affiliation{Universit\'e Lille 1,
Laboratoire PhLAM,
IRCICA, 59655 Villeneuve d'Ascq, France}
\author{G. Bouwmans}
\affiliation{Universit\'e Lille 1, Laboratoire PhLAM, IRCICA, 59655 Villeneuve d'Ascq, France}
\author{G. Martinelli}
\affiliation{Universit\'e Lille 1, Laboratoire PhLAM, IRCICA, 59655 Villeneuve d'Ascq, France}
\author{A. Mussot}
\affiliation{Universit\'e Lille 1, Laboratoire PhLAM, IRCICA, 59655 Villeneuve d'Ascq, France}

\date{\today}

\begin{abstract}
A simple analytical model is developed to analyze and explain the complex dynamics of the multi-peak modulation instability spectrum observed in dispersion oscillating optical fibers~[M. Droques \emph{et al.}, 37, 4832-4834 \emph{Opt. Lett.}, (2012)]. We provide a simple expression for the local parametric gain which shows that each of the multiple spectral components grows thanks to a quasi-phase-matching mechanism due to the periodicity of the waveguide parameters, in good agreement with numerical simulations and experiments. This simplified model is also successfully used to tailor the multi-peak modulation instability spectrum shape. These theoretical predictions are confirmed by experiments.
\end{abstract}

%Kerr effect, nonlinear optics, 42.65.Hw
%Frequency conversion (nonlinear optics), 42.65.Ky
%42.65.Wi 	Nonlinear waveguides
%Nonlinear guided waves, 42.65.Tg
%Nonlinear waveguides, optical, 42.65.Wi
%Four-wave mixing, 42.65.Hw

\maketitle

%%% environ 4000 mots %%%%%%%%%

\section{Introduction}

Modulation instability (MI) is a nonlinear process in which a weak perturbation is exponentially amplified by an intense field. MI has been investigated in many sub-fields of physics, especially in optics for the relative simplicity to perform experiments, in both homogeneous and periodical media~\cite{Abdullaev2002}. The additional degree of freedom brought by the periodicity has catched the attention of many research groups and has led to many theoretical and experimental works, in spatial~\cite{Meier2004,Centurion2006} and temporal Kerr media such as optical fibers~\cite{Murdoch1997,Murdoch1997a,Kikuchi1995,Shiraki1998}. This physical flexibility is of particular interest since the modulation period can be varied from the meter range up to tens of kilometers, leading to a large window of investigation. In the early nineties, the rise of optical telecommunication networks has led to the deployment of "natural" periodic optical fiber systems due to the alternation of all-optical regeneration devices and/or dispersion managed lines~\cite{Matera1993,Smith1996}. In addition to the fundamental interest brought by these systems, it was then necessary to understand in depth the origin of the characteristic spurious MI sidebands~\cite{Matera1993,Smith1996} which are highly detrimental for telecommunications since they are generated in the GHz range~\cite{Shiraki1998,Kikuchi1995} due to their period in the range of kilometers. A lot of theoretical studies have therefore been initiated in this context~\cite{Matera1993,Kikuchi1995,Smith1996,Bronski1996,Abdullaev1996,Kaewplung2002,Consolandi2002,Kumar2003,Ambomo2008}. %However, most of these results were quite complex to interpret and the physical insight quite difficult to understand for non-specialists.

Very recently, dispersion management was pushed one step further with the experimental demonstration of MI in continuously modulated waveguides~\cite{Driscoll2012,Droques2012}. From these results, the spurious consequence of the MI sidebands due to the periodicity can be turned into benefits since these works show the possibility to obtain multiple parametric gain bands in the THz range. It should then provide another degree of freedom for designing optical systems requiring broad bandwidths such as in all optical signal processing systems where there is a growing demand.

While a deep theoretical study of MI in periodically tapered fibers has been reported very recently~\citep{Armaroli2012}, we propose in the present work a simplified analytical treatment allowing to accurately describe the MI dynamics and to tailor the overall shape of its multi-peak spectrum. Besides providing insight into the underlying physics, our analytical treatment allows to derive an expression for the local linear parametric gain for the first time. Finally, in order to illustrate the practical interest of this simple analysis, we report experiments in which the multi-peak MI spectrum has been tailored to suppress one MI sideband or to favor a single strong one, in good agreement with our analytical predictions.

\section{Context}

\begin{figure}[t!]
\centering \includegraphics[width=8.4cm]{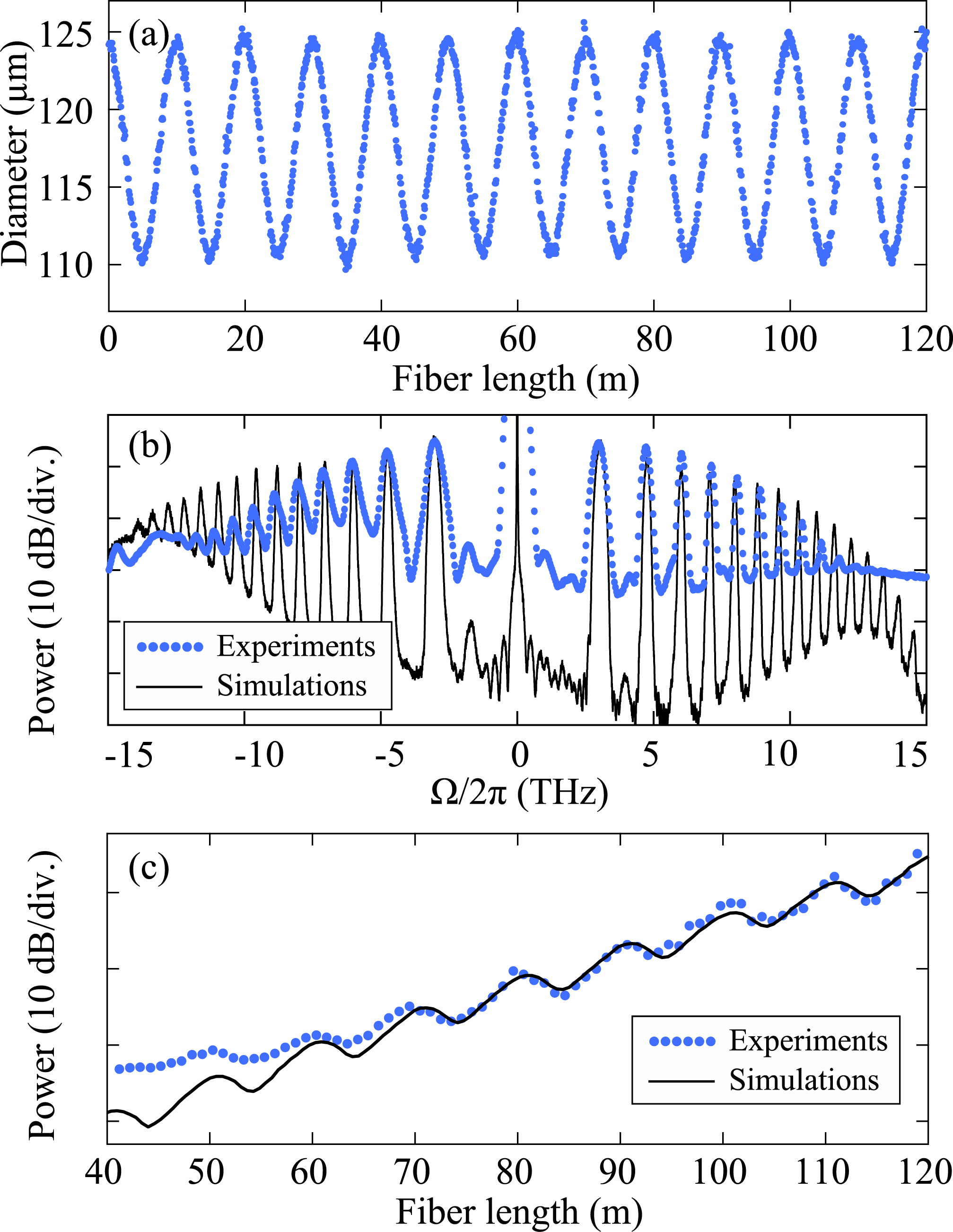} % high resolution JPEG ou PNG
      \caption{(a) Outer diameter of the DOF versus length measured during fiber drawing. (b) Experimental (circles) and simulated (solid line) spectra obtained for a pump power of 20~W and a fiber length of 120~m. (c) Evolution of the power of the first sideband ($k = 1$) versus fiber length obtained from experiments (circles) and numerical simulations (solid line). Results are from~\cite{Droques2012}.}
	  \label{fig1}
\end{figure}

We recently reported the first striking experimental demonstration of MI in a dispersion oscillating fiber (DOF)~\cite{Droques2012}. Although details can be found in~\cite{Droques2012}, the aim of this section is to briefly summarize our previous results in order to facilitate the reading of the present paper. Figure~\ref{fig1} shows a measurement of the evolution of the fiber diameter along its length. The outer diameter follows a sine shape with a modulation amplitude of $\pm7~\%$ and a period $Z$ of 10~m, which results in a dispersion modulation over the fiber length $z$ with a quasi-sinusoidal shape over the wavelength range of interest here, written as
\begin{equation}
\beta_2(z) = \overline{\beta_2}+\beta_2^{\mathrm{A}}\sin\left(\frac{2\pi z}{Z}\right) \label{eqbeta2}
\end{equation}
where $\overline{\beta_2} = 1.2\times 10^{-27}$~s$^2$/m is the average second order dispersion and $\beta_2^{\mathrm{A}} = 1.5\times10^{-27}$~s$^2$/m is the amplitude modulation at our pump wavelength of 1072~nm.

Full circles in Fig.~\ref{fig1}(b) correspond to the spontaneous MI spectrum obtained by pumping a 120~m-long sample of this DOF (labeled DOF\#1) with 2~ns pulses with a peak power $P_p$ of 20~W at $\lambda_p = 1072$~nm. The solid line represents the spectrum resulting from the numerical integration of the generalized nonlinear Schr\"{o}dinger equation (GNLSE) seeded by noise to accurately reproduce experimental random initial conditions (all details and parameters are given in~\cite{Droques2012}). Both spectra are in good agreement (except for the higher experimental noise floor) and show the generation of multiple MI sidebands pairs~\cite{kelly} spanning over more than 10~THz. The frequency of these parametric sidebands can be roughly estimated from a quasi-phase-matching relation developed in the case of an infinitely long grating~\cite{Matera1993,Kikuchi1995,Smith1996}
\begin{equation}
\overline{\beta_2}\Omega_k^2+2\gamma P_p=2\pi k/Z \label{eqPM}
\end{equation}
where $k$ is an integer, $\Omega_k$ is the pulsation detuning from the pump and $\gamma$ is the average nonlinear coefficient of the DOF.

The dynamics of the MI process with fiber length was investigated by cutting back the DOF and recording output spectra. As an illustration, Fig.~\ref{fig1}(c) shows the evolution of the power of the first sideband ($k = 1$) versus fiber length obtained from experiments (circles) and numerical simulations (solid line). The dynamics observed both in experiments and simulations exhibits periodic regions of deamplification, which makes the side lobe power oscillate along the fiber around the exponential growth (expected for a perfectly phase-matched process). Such a dynamics is expected from quasi-phase-matched processes, but it differs from the one observed in second-order nonlinear crystals in which there are no regions of deamplification. This particular and unusual feature will now be studied in detail with help of a simple and intuitive analytical model, in order to provide further insight into the underlying physics.

\section{Analytical model}\label{model}

\subsection{Parametric gain calculation}

It is well established that the MI process can be interpreted in the spectral domain as a four-wave mixing (FWM) process~\cite{Golovchenko1994,Harvey2003}. In this frame, the parametric FWM gain spectrum can be obtained by studying the stability of the steady state solution against weak perturbations through a so-called linear stability analysis. In dispersion managed optical systems, this tool has allowed to analytically predict the complex multi-peak gain spectrum~\cite{Smith1996,Abdullaev1996}, but such an analysis does not provide any clear insight into the dynamics of the process nor any details about the fine evolution of the field over a single modulation period of the fiber. To this aim, we propose here a more intuitive explanation of the results from Ref.~\cite{Droques2012} by revisiting a simplified truncated three-wave model usually aimed at describing Fermi-Pasta-Ulam recurrence and fiber-optic parametric amplification~\cite{Trillo1991,Inoue2001,VanSimaeys2002}. This model allows to account for the relative phase variations between pump, signal and idler waves during propagation. In our work, it will be induced by the longitudinal variations of dispersion rather than pump depletion. Our starting point is the four coupled differential equations given by Eqs.~3 in Ref.~\cite{Inoue2001}. We neglect fiber loss, we assume that the pump remains undepleted and that signal and idler powers, $P_s$ and $P_i$, are much less than the pump power $P_p$ over the whole DOF length. It is then easy to show that this system reduces to the following equations
\begin{subequations}
\begin{eqnarray}
\frac{dP_s(\Omega,z)}{dz}=2\gamma P_p \sqrt{P_s(\Omega,z)P_i(\Omega,z)} \sin\theta(\Omega,z)     \label{eq2a}\\
\frac{dP_i(\Omega,z)}{dz}=2\gamma P_p \sqrt{P_i(\Omega,z)P_s(\Omega,z)} \sin\theta(\Omega,z)	 \label{eq2b}  \\
\frac{d\theta(\Omega,z)}{dz}=\Omega^2\left[\overline{\beta_2}+\beta_2^{\mathrm{A}}\sin\left(\frac{2\pi z}{Z}\right) \right] + 2\gamma P_p \big(1 + \cos\left[\theta (\Omega,z)\right]\big)	\label{eq2c}
\end{eqnarray}\label{eq2}
\end{subequations}
where $\Omega$ is the shift of the signal and idler pulsations from the pump, and $\theta(\Omega,z)$ describes the longitudinal evolution of the relative phase difference between all these waves~\cite{Inoue2001}. The discrepancy between solutions of Eq.~\ref{eqPM} and experimental/numerical values mentioned in Ref.~\cite{Droques2012} can now be understood from Eq.~\ref{eq2c}. Indeed, Eq.~\ref{eqPM} assumes that the nonlinear phase mismatch can be approximated by $2\gamma P_p$~\cite{Matera1993,Kikuchi1995,Smith1996}, while Eq.~\ref{eq2c} shows that it is in fact equal to $2\gamma P_p\big(1+\cos\left[\theta(\Omega,z)\right]\big)$. This does not impact the validity of the present results since this additional term remain low for the pump powers involved in the present study. In order to obtain a simple analytic solution of the set of Eqs.~\ref{eq2}, we thus neglect the last term $\cos\left[\theta (\Omega,z)\right]$ in Eq.~\ref{eq2c}. It physically means that we assume that the longitudinal evolution of the nonlinear phase mismatch term is weak as compared to the linear and uniform nonlinear phase mismatch terms, which is valid for low pump powers. By integrating the set of Eqs.~\ref{eq2}, we find that the total accumulated gain of the signal in power writes as
\begin{equation}
G(\Omega,z)=\frac{P_s(\Omega,z)}{P_s(\Omega,0)} = \frac{1}{4}[1-\rho]+\frac{1}{4} [1+\rho +2\sqrt{\rho}] \mathrm{exp} \left[\int_0^z g(\Omega,z')dz'\right] \label{eqGain}
\end{equation}
with $\rho = P_i(\Omega,0)/P_s(\Omega,0)$. In the following, we set $\rho = 1$ for the sake of simplicity. In Eq.~\ref{eqGain}, $g(\Omega,z) = 2\gamma P_p \sin[\theta(\Omega,z)]$ is the local linear gain. Its calculation requires integrating Eq.~\ref{eq2c} in order to evaluate $\theta(\Omega,z)$, which gives (under our assumptions)
\begin{equation}
\theta(\Omega,z)=[\overline{\beta_2}\Omega^2 +2\gamma P_p]z+ \frac{\beta_2^{\mathrm{A}}\Omega^2}{2\pi /Z} [1-\cos(2\pi z/Z)]+\theta(\Omega,0) \label{equ16}
\end{equation}
Finally, by using a Fourier series expansion to calculate the $\sin[\theta(\Omega,z)]$ term, we find that the local linear gain writes as
\begin{equation}
g(\Omega,z)= 2\gamma P_p\sum
\limits_{{q=-\infty}}^{q=+\infty}J_q\left(\frac{\beta_2^{\mathrm{A}}\Omega^2}{2\pi/Z}\right) \sin\left[\left(\overline{\beta_2}\Omega^2+2\gamma P_p-\frac{q2\pi}{Z}\right)z+K_q\right]  \label{eqg}
\end{equation}
with $K_q=\frac{\beta_2^{\mathrm{A}}\Omega^2}{2\pi /Z}-q\frac{\pi}{2}+\theta(\Omega ,0)$. Thus, Eq.~\ref{eqg} indicates that the linear gain $g(\Omega,z)$ at a fixed pulsation detuning $\Omega$ can be interpreted as the sum of sine functions in $z$. These sine functions all have a zero average value except when their argument becomes independent of $z$. It only occurs at specific spectral components $\Omega$ (equal to the pulsation $\Omega_k$ in Eq.~\ref{eqPM}) corresponding to solutions of the quasi-phase-matching relation~\ref{eqPM}. For these specific pulsation detunings $\Omega_k$, each term of the sum in Eq.~(\ref{eqg}) leads to periodical amplification and deamplification phases along the DOF except for the uniform contribution corresponding to $q = k$. This last term therefore prevails over the other ones on the gain $G(\Omega,z)$ for long enough propagation distances. Thus the linear gain of the $k^{\mathrm{th}}$ spectral component mainly depends on this uniform term as long as the fiber exceeds a few modulation periods. It is then equal to $2\gamma P_p |J_{k}(\frac{\beta_2^{\mathrm{A}}\Omega_{k}^2}{2\pi/Z})|$, by choosing $K_k=+\frac{\pi}{2}$ as initial condition. Note that this is analogous to the choice of maximizing the gain in MI in uniform fibers by setting the initial phase mismatch value to $\pi/2$~\cite{Inoue2001, VanSimaeys2002}.

\begin{figure}[t!]
 \centering \includegraphics[width=8.4cm]{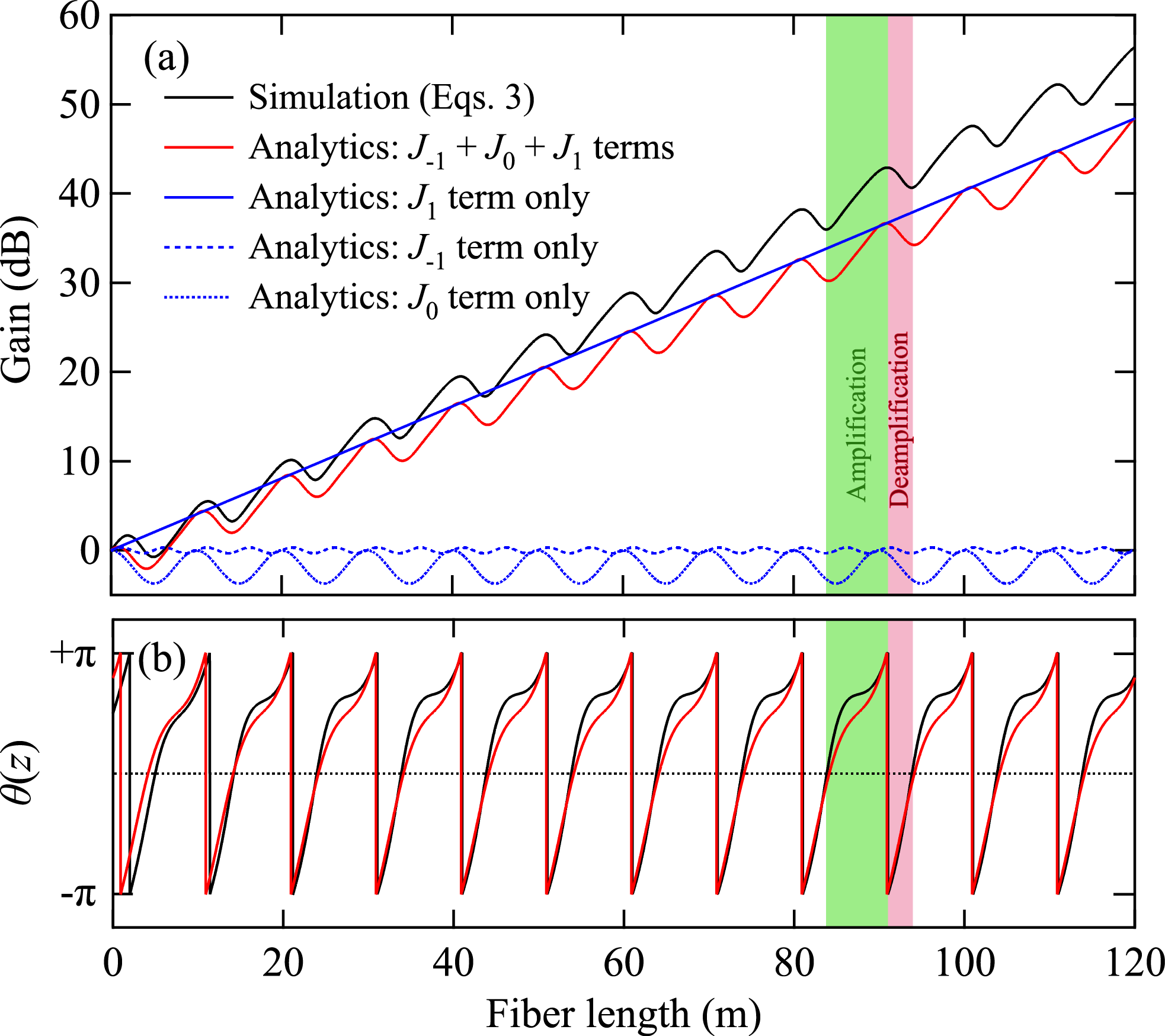} % high resolution JPEG ou PNG
      \caption{(a) Evolution of the gain of the first amplified frequency ($k = 1$) from Eq.~\ref{eqGain}, in red
line with the contribution of $J_{1}$+$J_{0}$+$J_{-1}$ (average gain + oscillating terms), in blue solid line for $J_{1}$ only, in blue dotted line for $J_0$ only and in blue dashed line for $J_{-1}$ only with $\Omega_{\mathrm{max}}^{\mathrm{theo}}=2\pi\times 2.63\times10^{12}$ rad/s (Eq.~\ref{eqPM}). The solid black line is calculated from the numerical integration of the original set of equations (Eqs.~\ref{eq2}) with $\Omega_{\mathrm{max}}^{\mathrm{simu}}=2\pi\times 2.93\times10^{12}$ rad/s. (b) Evolution of $\theta(z)$ from our analytical study (solid red line) and from numerical simulations with Eqs.~\ref{eq2} (solid black line). The green area corresponds to amplification and the red one to deamplification over one period.}
	  \label{fig2}
\end{figure}
\begin{figure}[t!]
 \centering \includegraphics[width=8.4cm]{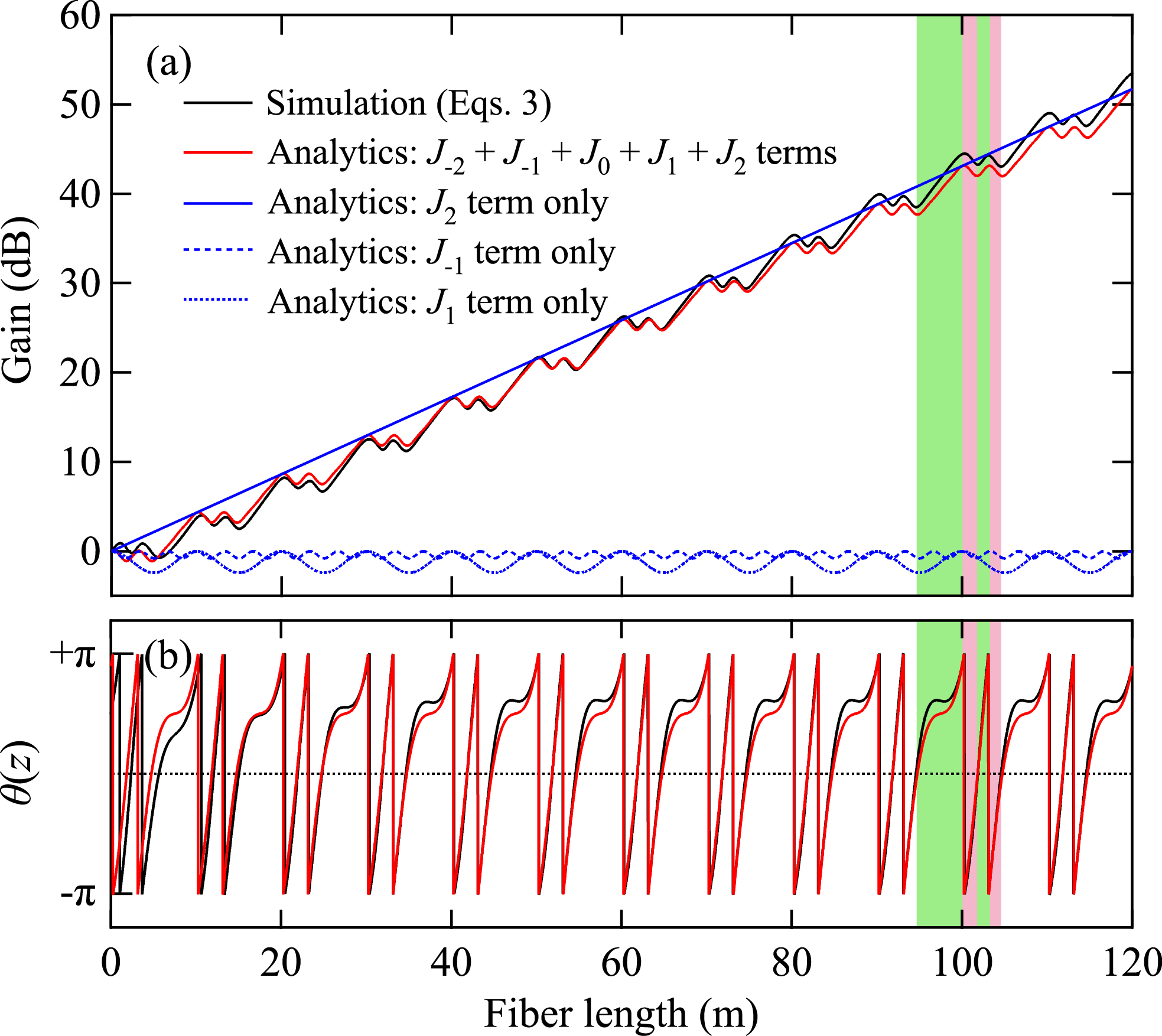} % high resolution JPEG ou PNG
      \caption{(a) Evolution of the gain of the second amplified frequency ($k = 2$) from Eq.~\ref{eqGain}, in red
line with the contribution of $J_{-2}$+$J_{-1}$+$J_{0}$+$J_{1}$+$J_{2}$ (average gain + oscillating terms), in blue solid line for $J_{2}$ only, in blue dotted line for $J_1$ only and in blue dashed line for $J_{-1}$ only with $\Omega_{\mathrm{max}}^{\mathrm{theo}}=2\pi\times 4.49\times10^{12}$ rad/s (Eq.~\ref{eqPM}). The solid black line is calculated from the numerical integration of the original set of equations (Eqs.~\ref{eq2}) with $\Omega_{\mathrm{max}}^{\mathrm{simu}}=2\pi\times 4.61\times10^{12}$ rad/s. (b) Evolution of $\theta(z)$ from our analytical study (solid red line) and from numerical simulations with Eqs.~\ref{eq2} (solid black line). Green areas correspond to amplification and the red ones to deamplification over one period.}
	  \label{fig3}
\end{figure}

\subsection{Physical interpretation}

To illustrate this process, we firstly focus on the first spectral component ($k = 1$). The solid black line in Fig.~\ref{fig2}(a) shows the evolution of the maximum gain (at $\Omega = \Omega_{\mathrm{max}}^{\mathrm{simu}}$) obtained from numerical integration of the complete set of original Eqs.~\ref{eq2}. Note that an excellent agreement is achieved with the numerical integration of the GNLSE (not shown here for the sake of clarity). The blue solid line in Fig.~\ref{fig2}(a) corresponds to the term of uniform gain (Bessel function $J_{1}$), the blue dotted and dashed lines correspond to the highest amplitude oscillating terms (Bessel functions $J_{0}$ and $J_{-1}$ respectively in this case) and the red solid line corresponds to their sum. We limit our investigations to $J_{0}$ and $J_{-1}$ because all other Bessel functions have much lower contributions in this example. A good agreement is obtained between the red solid curve from the analytical model, and the black one from numerical simulations, which confirms the validity of our assumptions and the accuracy of our method. In each modulation period, the amplification phase is characterized by $0<\theta(\Omega,z)<\pi$ and the deamplification one has $-\pi<\theta (\Omega,z)<0$ as represented in Fig.~\ref{fig2}(b), the total phase shift being equal to $2\pi$ per period.

The dynamics of the second spectral component ($k = 2$) is shown Figure~\ref{fig3}. The same reasoning as for the first one ($k = 1$) can be applied. The $J_{2}$ term provides the average exponential gain (blue solid line) and additional oscillating terms provide the oscillating behavior of the overall gain. By taking the five terms with highest amplitude into account (from $k = -2$ to $k = 2$) a good agreement between Eq.~\ref{eqGain} and numerical simulations from Eqs.~\ref{eq2} is achieved. Note that adding other higher-order terms does not significantly change analytical results (displayed in red curves). For the sake of clarity, only the two highest amplitude ones ($J_{-1}$ and $J_{1}$) are represented in Fig.~\ref{fig3}(a) (in dashed and dotted lines respectively). In this case, there are two amplification and deamplification phases per period. The evolution of the phase represented in Fig.~\ref{fig3}(b) also shows a more complex evolution than for the first ($k = 1$) sideband and the total phase shift is now equal to $4\pi$ per period. The agreement between numerical simulations from Eqs.~\ref{eq2} (black lines) and our analytical result (red lines) is here excellent both for the evolution of the gain and of the phase.
%, which further illustrates the quasi-phase-matching mechanisms involved in the process.

Figures~\ref{fig2} and \ref{fig3} emphasize that the dispersion modulation enables to control the evolution of the relative phase of the waves so that the whole process can be seen as quasi-phase-matched, the variation of the relative phase over one period being equal to $2k\pi$ for the $k^{\mathrm{th}}$ spectral component. Indeed, this relative phase $\theta(\Omega,z)$ would grow linearly in the absence of the modulation term ($\beta_2^{\mathrm{A}}=0$) in Eq.~\ref{eq2c}~\cite{gain}. This linear growth would lead to amplification and deamplification phases of same length and consequently the total accumulated gain would be negligible.

From a more practical point of view, the frequency of the spectral component $\Omega_k$ can be widely modified simply by changing the periodicity of the grating (as in a diffraction grating for the position of its different orders), while the gain (analogous to the diffraction efficiency in a specific order) can be modulated independently through the ratio $\beta_2^{\mathrm{A}}/\overline{\beta_2}$. Note however that the deamplification phases of the signal along the fiber cannot be totally avoided but only reduced. This can be understood either by considering that they are due to the contribution of all oscillating terms of Eq.~\ref{eqg} which cannot be all suppressed simultaneously, or by considering that the dispersion grating enables the evolution of $\theta(\Omega,z)$ to deviate from a linear growth but not to limit its evolution in the $[0;\pi]$ range required for a positive gain (see Eq.~\ref{eq2a}).

\section{Evolution of the MI spectrum with average dispersion}\label{dyn}

In order to further emphasize the accuracy of our simplified analytical model, we studied the evolution of the MI spectrum as a function of average dispersion value at the pump frequency both in normal and anomalous dispersion regimes. We chose the DOF parameters so that they match the ones of the fiber used hereafter. We took into account longitudinal variations of $\beta_2$ according to Eq.~\ref{eqbeta2}, and average values of $\gamma$, $\beta_3$ and $\beta_4$ instead of their longitudinal evolution. We checked numerically that this has negligible impact in our conditions. The DOF parameters are $\beta_2^{\mathrm{A}}$ of $10^{-27}$~s$^2$/m, $\overline{\beta_3} = 6.8\times10^{-41}$~s$^3$/m, $\overline{\beta_4} = 1.7\times10^{-55}$~s$^4$/m and $\gamma =7$~W$^{-1}$.km$^{-1}$ at 1064~nm. The DOF is 120~m long and the modulation period is $Z = 10$~m. Fiber attenuation and stimulated Raman scattering are neglected.

We have firstly performed numerical simulations using Eqs.~\ref{eqGain} by varying values of $\overline{\beta_2}$ from -1.5 to 1.5$\times$10$^{-27}$~s$^2$/m.
\begin{figure}[t!]
 \centering \includegraphics[width=12cm]{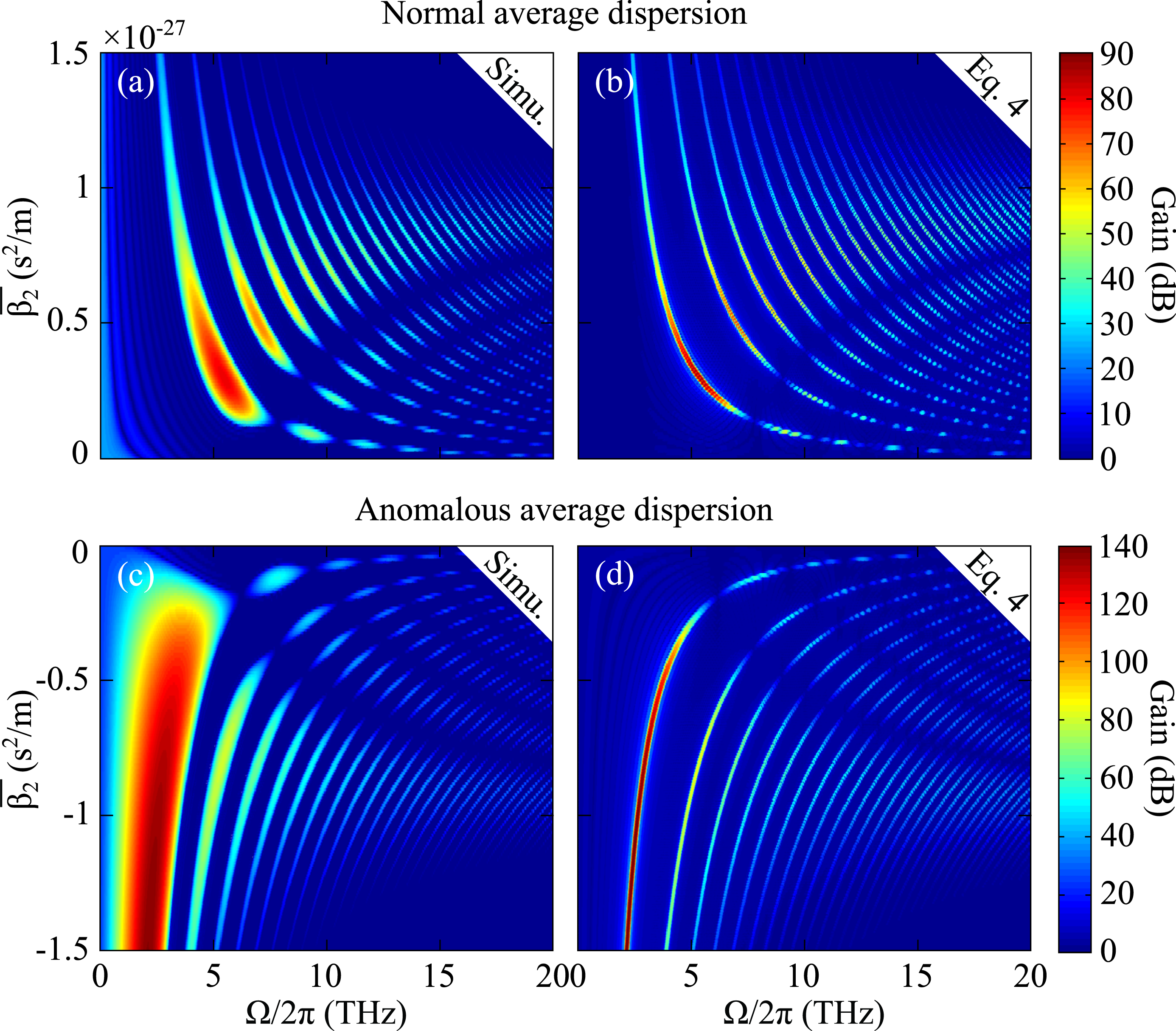} % high resolution JPEG ou PNG
      \caption{Top view of the gain spectrum (in logarithmic scale) obtained from (a),(c) GNLSE simulations and from (b),(d) our analytical model (Eq.~\ref{eqGain}) as a function of the average second-order dispersion. (a),(b) panels corresponds to average normal dispersion pumping and (c),(d) to anomalous average dispersion pumping. The zero frequency corresponds to the pump.}
	  \label{fig4}
\end{figure}
The maps displayed in Fig.~\ref{fig4}(a) and (c) represent calculated numerical gain spectra (in logarithmic scale) for varying $\overline{\beta_2}$ values in normal and anomalous dispersion regions respectively. %Note that the gain color code is different for both average dispersion regimes because the gain of classical MI sidebands in anomalous dispersion (corresponding to $k = 0$) is much higher than the one obtained thanks to the periodicity ( corresponding to other $k$ values) in this example.
For the sake of clarity, only one half of the overall spectrum is displayed, the other half being perfectly symmetric with respect to the pump frequency. In parallel, we have plotted in Fig.~\ref{fig4}(b) and (d) the same map using the analytical gain expression given by Eq.~\ref{eqGain} with the same parameters as for the numerical simulations. There is a good qualitative agreement between both graphs, although quantitative agreement is not reached due to our approximations explained above. However, our model allows to reproduce the specific dynamical features observed in numerical simulations:
\begin{itemize}
\item Firstly, the detuning of each MI sideband from the pump decreases as $\overline{\beta_2}$ increases, as expected from Eq.~\ref{eqPM}.
\item Secondly, the gain strongly depends on $\overline{\beta_2}$, which differs from classical MI spectrum observed in uniform fibers in the anomalous dispersion regime, in which the maximal parametric gain does not depend on $\beta_2$. Additionally, the maximal gain does not correspond to the same $\overline{\beta_2}$ value for each MI sideband.
\item Thirdly, we can identify specific $\overline{\beta_2}$ values for which one or several initially well-defined MI sideband are cancelled, i.e. their parametric gain vanishes. The frequency of cancelled sidebands increases with increasing $\overline{\beta_2}$ values, for each value of $k$. This \emph{a priori} unexpected cancellation of parametric gain is in fact due to vanishing $J_k$ functions in Eq~\ref{eqg} and will be detailed in the last section of this paper.
\end{itemize}

\section{Tailoring the MI gain spectrum}
The simple analytical approach presented in section \ref{model} and further confirmed in section \ref{dyn} allows a better understanding of the complex dynamics of the process and it has allowed us to design experiments in which the multi-peak gain spectrum is tailored. To illustrate this, we focus our attention here to two striking examples. We chose either to completely cancel a given spectral component (as previously mentioned in section \ref{dyn}) or to maximize the gain of a sideband pair with regards to the others. To reach these goals, let us recall that, as detailed in the analytical model above, the linear gain of the $k^{\mathrm{th}}$ spectral component can be approximated by
\begin{equation}
g(\Omega_k,z)=  2\gamma P_p \left|J_{k}\left(\frac{\beta_2^{\mathrm{A}}\Omega^2_k}{2\pi/Z}\right)\right|=2\gamma P_p \left|J_{k}\left[\frac{\beta_2^{\mathrm{A}}}{\overline{\beta_2}}\left(k-\frac{\gamma P_p Z}{\pi}\right)\right]\right|
\label{eqJ}
\end{equation}
Equation~\ref{eqJ} indicates that the gain of the $k^{\mathrm{th}}$ spectral component can be totally cancelled by simply finding the argument $\eta = \frac{\beta_2^{\mathrm{A}}\Omega_k^2}{2\pi/Z} = \frac{\beta_2^{\mathrm{A}}}{\overline{\beta_2}}\left(k-\frac{\gamma P_p Z}{\pi}\right)$ for which the Bessel function $J_{k}$ vanishes. $\eta$ can be adjusted by controlling the modulation amplitude of dispersion, $\beta_2^{\mathrm{A}}$, or the fiber period, $Z$, which both require manufacturing new DOF samples. But it can also be adjusted by controlling the average dispersion at the pump wavelength, $\overline{\beta_2}$, which can be done experimentally by simply tuning the pump wavelength.

\subsection{Cancellation of spectral component}

\begin{figure}[t!]
 \centering \includegraphics[width=8.4cm]{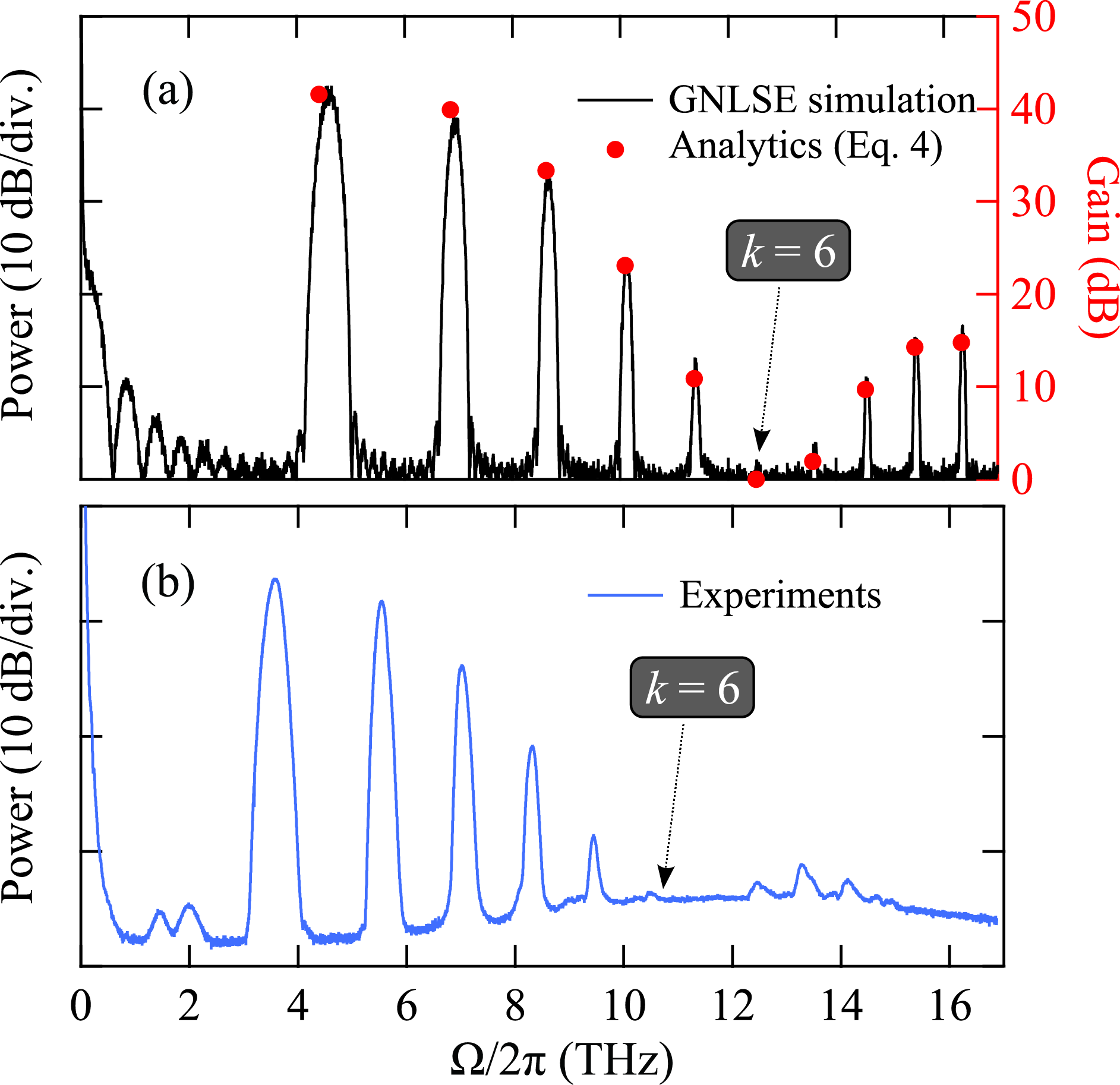} % high resolution JPEG ou PNG
      \caption{Illustration of the cancellation of the $k = 6$ spectral component. (a) Maximal gain obtained from Eq.~\ref{eqJ} (circles, right axis) and output spectrum simulated with the GNLSE (solid line, left axis), for $\overline{\beta_2} = 5.8\times10^{-28}$~s$^2$/m and $P_p = 13$~W. (b) Corresponding experiments performed in DOF\#2 for a pump wavelength of 1067.5~nm and pump power of 24~W.}
      \label{fig5}
\end{figure}

To illustrate this, we fabricated a new DOF sample, labeled DOF\#2 hereafter. DOF\#2 is 120~m-long and it has an average dispersion $\overline{\beta_2}$ of $10^{-27}$~s$^2$/m, a modulation amplitude $\beta_2^{\mathrm{A}}$ of $10^{-27}$~s$^2$/m at 1064~nm and its ZDW oscillates between 1064~nm and 1080~nm. The average third order dispersion term is $\beta_3 = 6.8\times10^{-41}$~s$^3$/m, the average nonlinear coefficient is $\gamma =7$~W$^{-1}$.km$^{-1}$ and the attenuation is $\alpha = 7.5$~dB/km at 1064~nm. To illustrate the cancellation of parametric gain at specific frequencies, we choose for example to cancel the $k = 6$ sideband pair. In this case, we found that $J_6$ vanishes for a $\overline{\beta_2}$ of $5.8\times10^{-28}$~s$^2$/m with the parameters of DOF\#2 given above and a pump power of 13~W. Circles in Fig.~\ref{fig5}(a) shows the gain calculated with the above model (Eq.~\ref{eqJ}) for each spectral component and for a fiber length of 120~m, while the solid line represents the output spectrum obtained from a numerical integration of the GNLSE with a pump power of 13~W. We average 50 output spectra seeded by random initial conditions to account for the averaging performed during the experimental recording of a spectrum. These results firstly confirms the ability of our simplified model to correctly predict the maximal gain of each sideband and they also show that the $k = 6$ spectral component is indeed canceled. Experiments performed in DOF\#2 by tuning the pump wavelength to 1067.5~nm (which is close to the required $\overline{\beta_2}$ value of $5.8\times10^{-28}$~s$^2$/m) are displayed in Fig.~\ref{fig5}(b).
The overall shape of the experimental spectrum nicely matches the one obtained from theory and this measurement also confirms the cancellation of the $6^{\mathrm{th}}$ peak. In all experiments presented in this section, the pump power was the only adjustable parameter. It had to be increased up to 24~W to observe the expected behaviors, which is higher than the power of 13~W used in simulations and in the model. This discrepancy in pump power is reasonable given the uncertainty on the evaluation of fiber properties (attenuation, dispersion and nonlinearity) and on the pump laser parameters (repetition rate, pulse duration, measurement of average power).

\subsection{Maximization of a single spectral component}

\begin{figure}[t!]
 \centering \includegraphics[width=8.4cm]{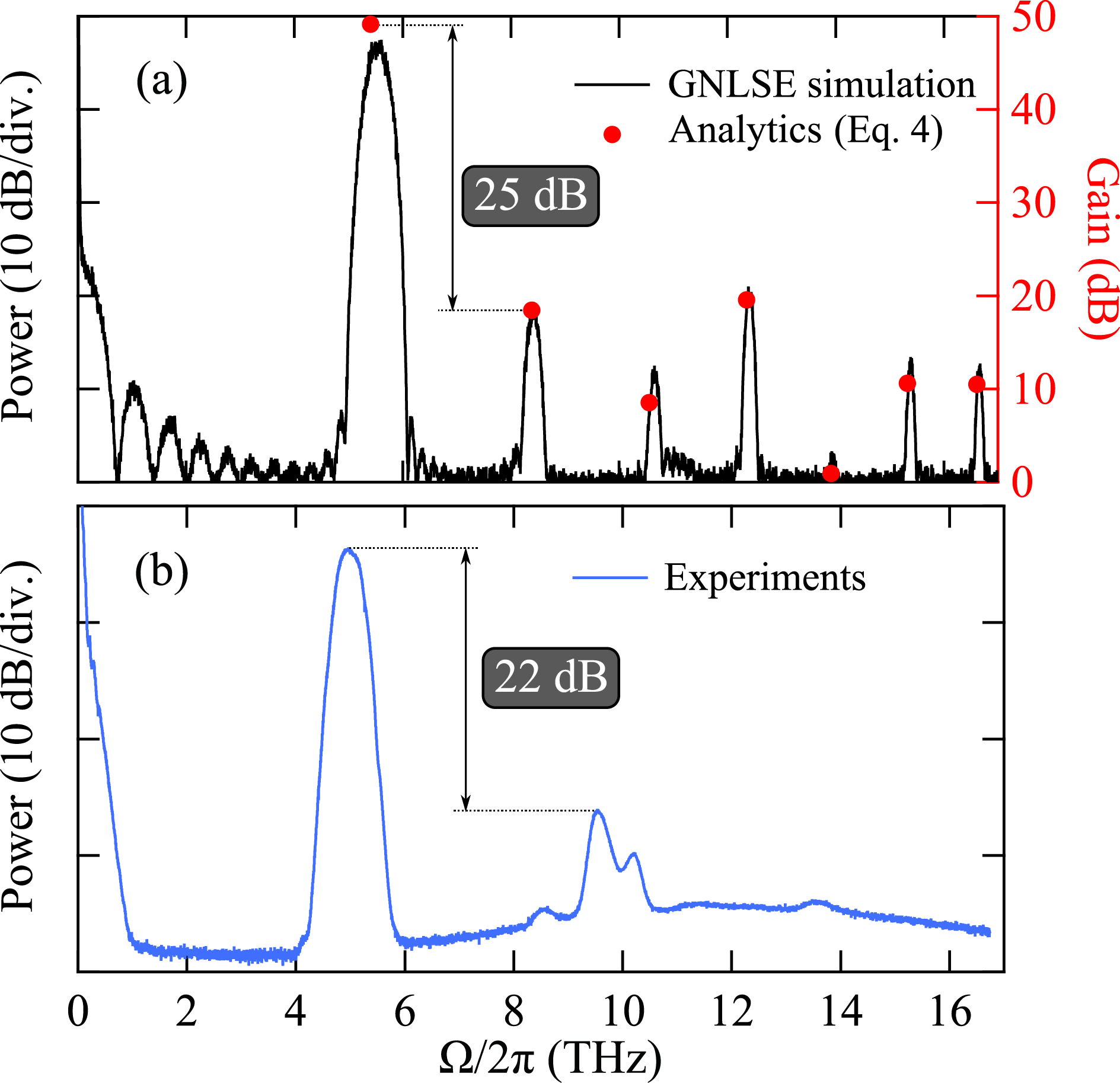} % high resolution JPEG ou PNG
      \caption{Illustration of the maximization of the $k = 1$ spectral component. (a) Maximal gain obtained from Eq.~\ref{eqJ} (circles, right axis) and simulated output spectrum (solid line, left axis), for $\overline{\beta_2} = 3.87\times10^{-28}$~s$^2$/m and $P_p = 13$~W. (b) Corresponding experiments performed in DOF\#2 for a pump wavelength of 1071.5~nm and pump power of 24~W.}
      \label{fig6}
\end{figure}

In order to further illustrate the possibility of tailoring the multi-peaks MI spectrum, we used Eq.~\ref{eqJ} to find a configuration in which the $k = 1$ sideband is maximized, i.e. it experiences a much higher gain than any other ones. In this case, we simply need to find a $\overline{\beta_2}$ value (and thus a value of the $\eta$ parameter), experimentally a pump wavelength,  which maximizes the $J_1$ Bessel function. Figure~\ref{fig6}(a) shows the gain calculated from Eq.~\ref{eqJ} in circles, as well as the output spectrum obtained from numerical integration of the GNLSE for a $\overline{\beta_2}$ value of $3.87\times10^{-28}$~s$^2$/m. These results are again in excellent agreement, and they indeed show that the first sidelobe is favored since it has a 25~dB gain higher than all other ones. Experiments were performed by accordingly tuning the pump wavelength to 1071.5~nm. The output spectrum plotted in Fig.~\ref{fig6}(b) shows that the power of the first sidelobe is 22~dB higher than other spectral components, in good agreement with theoretical predictions.

It is also worth noting that in this case, the argument of the $J_1$ Bessel function has been chosen so that the gain value of the first sideband calculated from our model corresponds to exactly the maximum value of 0.582 for $J_1$. This means that the gain brought by the periodicity for this sideband approximately equals 0.582$\times 2\gamma P_p$ according to Eq.~\ref{eqJ}. It is less than a factor of 2 lower than the maximal gain expected from a classical MI process in the anomalous dispersion region in uniform fibers (which would approximately be equal to $2\gamma P_p$). This observation is all the more important given that no gain is expected in the normal dispersion region in uniform fibers (neglecting higher-order dispersion terms and higher-order fiber modes).

Although the control of the overall MI spectrum shape requires a change of quasi-phase-matched frequencies, these examples demonstrate the possibility of harnessing the MI spectrum thanks to the periodic dispersion landscape. A simultaneous control of both the spectral shape and sideband frequencies would still be possible by simultaneously adjusting $\overline{\beta_2}$ and $\beta_2^{\mathrm{A}}$, which would however require manufacturing new fibers.

\section{Conclusion}

Following our first experimental demonstration of MI in dispersion oscillating fibers~\cite{Droques2012}, we have investigated this process theoretically here. Starting from the well-known truncated three-wave model, we have derived an approximate analytical expression to predict the local parametric gain. This simplified model gives good agreement with numerical simulations and experiments. It has also allowed to interpret the MI process in terms of a quasi-phase-matching mechanism due to the periodic nature of the fiber dispersion landscape. We have also used this model to emphasize the possibility of tailoring the MI spectrum, which has been confirmed experimentally by the cancellation or maximization of chosen spectral components.

Dispersion oscillating photonic crystal fibers such as the ones reported here and in~\cite{Droques2012} pave the way to a range of linear and nonlinear guided wave optical processes thanks to the longitudinal periodic modulation of their waveguiding properties. They should find applications in wavelength conversion, parametric amplification, generation of ultra-short pulse trains or soliton management among others.

% If you have acknowledgments, this puts in the proper section head.

\section*{Acknowledgments}
This work was partly supported by the ANR (IMFINI project), by the French Ministry of Higher Education and Research, the Nord-Pas de Calais Regional Council and FEDER through the "Contrat de Projets Etat R\'{e}gion (CPER) 2007-2013" and the "Campus Intelligence Ambiante" (CIA).

% Create the reference section using BibTeX:
\bibliographystyle{osa}
%\bibliography{biblio}

\begin{thebibliography}{10}
\bibitem{gain}
This is due to the fact that the non uniform nonlinear phase mismatch term
  $\cos\left[\theta (\Omega,z)\right]$ in Eq.~\ref{eq2c} is neglected in this
  section.
\end{thebibliography}
%

\end{document}